\documentclass[10pt,twocolumn]{revtex4}

\usepackage{graphicx}
\usepackage{dcolumn}
\usepackage{bm}
\usepackage{float}
\usepackage{placeins}

\begin{document}

\title[Coinfection outbreaks in temporal networks]{Risk of coinfection outbreaks in temporal networks: a case study of a hospital contact network} 
\author{Jorge P. Rodr\'iguez$^{1}$}
\author{Fakhteh Ghanbarnejad$^{2}$}
\email[]{fakhteh.ghanbarnejad@gmail.com}
\author{V\'ictor M. Egu\'iluz$^{1}$}
\email[]{victor@ifisc.uib-csic.es}
\affiliation{$^{1}$Instituto de F\'isica Interdisciplinar y Sistemas Complejos IFISC, CSIC-UIB, Palma de Mallorca, Spain \\
$^{2}$Institut f\"ur Theoretische Physik, Technische Universit\"at Berlin, Berlin, Germany}
\begin{abstract}
We study the spreading of cooperative infections in an empirical temporal network of contacts between people, including health care workers and patients, in a hospital. The system exhibits a phase transition leading to one or several endemic branches, depending on the connectivity pattern and the temporal correlations. There are two endemic branches in the original setting and the non-cooperative case. However, the cooperative interaction between infections reinforces the upper branch, leading to a smaller epidemic threshold and a higher probability for having a big outbreak. We show the microscopic mechanisms leading to these differences, characterize three different risks, and use the influenza features as an example for this dynamics.

\end{abstract}
\maketitle

\section{Introduction}
Infectious diseases have been a serious problem across the whole history of humankind. Nowadays, 200,000 hospitalizations are directly associated with influenza every year in the United States \cite{thompson2003mortality,thompson2004influenza}, and a potential pandemic could kill between 50 and 80 million people through a virulence strain similar to the 1918 influenza \cite{murray2007estimation,parrish2005origins}. However, the complexity of this dynamics is even higher, as infections can interact between themselves in several ways, inducing higher susceptibility or cross-immunity \cite{recker2007generation,newman2005threshold,karrer2011competing,sanz2014dynamics}. Hospitals, where different diseases are more likely to meet, are risky places for coinfection, that is the concurrent infection of a host with multiple pathogens. Moreover, diseases in hospitals are probably in their most virulent stages, weakening their hosts' immune system such that secondary infections are very likely to occur. Indeed, hospital acquired infections affect, on average, to 10\% of the admitted patients \cite{filetoth2008hospital}, while the network defined by the transfer of patients across hospitals may explain the spread of bacterial infections \cite{gracia2015spread}. Coinfection in hospitals has been reported, like the association between \emph{Clostridium difficile} and vancomycin-resistant \emph{Enterococcus} \cite{poduval2000clostridium}.  Some studies show that multiple pathogen infections were present in the children hospitalized \cite{nascimento2010,mansbach2012,bonzel2008}. P\'erez-Garc\'ia \emph{et al.} also reported that co-infection was significantly associated with nosocomial acquisition \cite{perez2016}.  This is translated not only in health issues, but also economic costs \cite{plowman2001rate}. 

In this work, we address the study of contagion processes following the temporal interactions pattern within a hospital. This can help determining the main drivers of this complex interacting spreading process and then preventing these cases. Thus we are motivated to calculate risks of coinfection outbreaks in a hospital. 
  
Recently, a model for cooperation between two infections has been proposed, with both infections following a dynamics that is an extension of the usual SIR (Susceptible-Infected-Recovered) model \cite{kermack1927contribution}, but considering that individuals that previously suffered from one infection are more likely to get a second infection than susceptible ones, in a process that is called coinfection \cite{chen2013outbreaks}. In contrast with usual SIR model applied to a single disease spreading along a network, this dynamics leads to abrupt transitions in several topologies \cite{cai2015avalanche,grassberger2016phase}. We study this dynamics in an empirical temporal network, to assess the differences that cooperation between infections can induce in real world. In fact, several studies have shown the differences between temporal and static approaches when dynamical processes occur in networks, changing the characteristic time scales of the processes \cite{karsai2011small,holme2012temporal,scholtes2014causality,masuda2016accelerating}. The temporal correlations have a remarkable effect on cooperation in evolutionary dynamics \cite{cardillo2014evolutionary}, and the interplay of them with the structure strongly affects the diffusion processes \cite{delvenne2015diffusion}. Nowadays, empirical temporal contact networks in hospital are becoming available for scientific analysis. For instance, Jarynowski and Liljeros made a dataset based on the registry of visits in the hospital was used to specify the interactions between hosts \cite{jarynowski2015contact}. Also contact networks for disease spreading in hospitals have been studied \cite{obadia2015detailed}, where the sequence of contacts was aggregated in a daily scale. 

Considering these motivations, we apply our cooperative spreading model to an empirical network built from hospital contacts. Then we make another step further and calculate risks of coinfection outbreaks on this hospital contact network, for a general case and the specific case of influenza. The body of the paper is structured in 4 sections, including the description of the empirical network, the definition of the dynamics describing the spreading process, the introduction of the risks that will be assessed and, finally, the results and discussions.

\section{Network description}

The empirical network we study here includes the contacts between $N=75$ people in a hospital (46 health care workers and 29 patients) \cite{vanhems2013estimating}. Contacts are reported by tracking devices when two individuals are located within a distance of 1--1.5 m, and are agreggated in time windows of $\Delta t=20\text{ s}$. Specifically, 32424 contacts were recorded along 17375 intervals, $t_{\text{max}}=96.53\text{ h}$, which gives an average degree per unit time $8.96\text{ h}^{-1}$. This network is described with a temporal adjacency matrix $A_{ij} (t)$, where $A_{ij} (t)=1$ if individuals $i$ and $j$ are connected at time $t$ and $A_{ij}(t)=0$ otherwise. Note that $A_{ij} (t)=A_{ji}(t)$ due to the symmetry of interactions.

  \begin{figure}[h!]
  \begin{center}
  \includegraphics[width=0.5\textwidth]{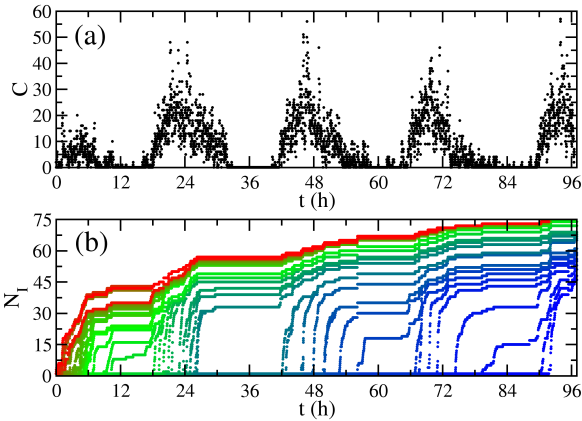}
  \caption{{\bf Temporal features of the network.}
(a) Number of contacts $C(t)=\sum\limits_{i,j>i} A_{ij}(t)$ happening at time $t$, aggregated for time intervals of 80 s. The low activity periods are valleys associated with night. (b) Number of infected individuals $N_I$ in a SI process with transmission probability equal to 1. Each curve starts from a different initially infected individual, assigning red color to trajectories whose initial infected individual is the first having a contact in the sequence, progressively changing to blue (last).}
      \label{fig1}
      \end{center}
      \end{figure}

This temporal network of contacts has a non-uniform activity on time, having peaks separated by $24\text{ h}$ (Fig. \ref{fig1}a). In fact, there are valleys of low activity, associated with night, such that dynamics starting here is very likely to die out due to the low number of contacts. This motivates our choice of the recovery probability according to the characteristic time scale for those valleys of 1000 time intervals ($\approx 5.5\text{ h}$). Speficically, in a SI (Susceptible-Infected) process with transmission probability equal to 1, where one disease is transmitted in every contact between an infected individual and a susceptible one, starting from one initially infected individual, the low activity periods have a constant number of infected individuals, until the morning arrives and new infections occur (Fig. \ref{fig1}b).

\section{Dynamics}

\begin{figure}[h!]
  \begin{center}
  \includegraphics[width=0.5\textwidth]{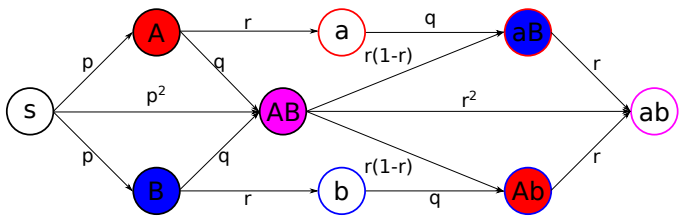}
  \caption{{\bf Scheme of the coinfection dynamics.}
      Probabilities of transition, under the exposure to the suitable active states, between different states are depicted in the center of the arrows connecting states. Primary infections, \textit{ i.e.} when a state S gets infected, happen with probability $p$, secondary infections, which are those from active/recovered for/from one infection to active for the other, happen with probability $q$, and recovery, from active to inactive states, happens with probability $r$. Filling colours indicate active states, while stroke colours stand for recovered states (red for infection $\mathcal{A}$, blue for infection $\mathcal{B}$, magenta for both). }
      \label{fig2}
      \end{center}
      \end{figure}

We consider a model where two infections, $\mathcal{A}$ and $\mathcal{B}$, both experiencing SIR dynamics, spread along the temporal contact network (Fig. \ref{fig2}). This leads to 9 different states for individuals:

{\bf Active states}
\begin{itemize}
\item $A$: singly infected with $\mathcal{A}$
\item $B$: singly infected with $\mathcal{B}$
\item $AB$: doubly infected with both $\mathcal{A}$ and $\mathcal{B}$
\item $aB$: recovered from $\mathcal{A}$, infected with $\mathcal{B}$
\item $Ab$: infected with $\mathcal{A}$, recovered from $\mathcal{B}$
\end{itemize}
{\bf Inactive states}
\begin{itemize}
\item $S$: susceptible
\item $a$: recovered from $\mathcal{A}$
\item $b$: recovered from $\mathcal{B}$
\item $ab$: recovered from both $\mathcal{A}$ anf $\mathcal{B}$
\end{itemize}

Indivuduals can get infected by their active neighbours. The states are updated synchronously according to the following rules:
\begin{enumerate}
\item Individuals that are active for $\mathcal{A}$ infect their susceptible neighbours with probability $p$ (analogous for individuals that are active for $\mathcal{B}$).
\item Secondary infections will occur when an individual with an active state is connected with an individual which has previously suffered from the other infection, happening with probability $q$.
\item Recovery from each infection in active individuals will happen independently with probability $r$.
\end{enumerate}

The case $q>p$ is representative for cooperation between infections, while $q=p$ means two independent spreading processes and $q<p$ would represent cross-immunity. 

As several individuals may remain active after the $t_{\text{max}}$ time steps of the empirical network, we repeat periodically the observed interaction pattern until there are no remaining individuals in active states. This means that, for time step $t$, we will consider the network of interactions $A_{ij} (t \bmod \  t_{\text{max}})$. Taking into account the analysis performed in Fig. \ref{fig1}, we set $r=0.001$, such that infections can survive the low activity night period.

\section{Estimation of risks}

We consider three estimators for the risk. First of all, the {\bf presence of an outbreak}: classical studies of disease spreading show a phase transition from non-endemic to endemic regime, when the transmission probability $p$ is changed; there is a threshold $p_c$, the critical point, such that for $p<p_c$ there are no outbreaks, while for $p>p_c$ the probability of having an outbreak grows. This epidemic threshold $p_c$ is calculated as the maximum of the susceptibility. Secondly, we will estimate the {\bf fraction of the population affected by the outbreak}: given a value for $p$, we compute the period prevalence \cite{rothman2012epidemiology}, defined in our study as the fraction of people $\rho_{ab}$ which has been coinfected in independent realizations in a statistical ensemble. Finally, we analize the distribution of $\rho_{ab}$, \emph{i.e.} the probability $\Pi_p(\rho_{ab})$ of having an outbreak of size $\rho_{ab}$ for a given $p$, that is useful to characterize the {\bf probability of having big outbreaks}, considering that a realization is a big outbreak when it is in the upper endemic branch on the $\rho_{ab}$-$p$ diagram. $\Pi_p(\rho_{ab})=0$ for $p<p_c$ and $\rho_{ab}>\frac{1}{N}$ (one initially doubly infected individual), and it will grow for values higher than $p_c$.

\section{Results}
\begin{figure*}[hbt!]
  \includegraphics[width=0.8\textwidth]{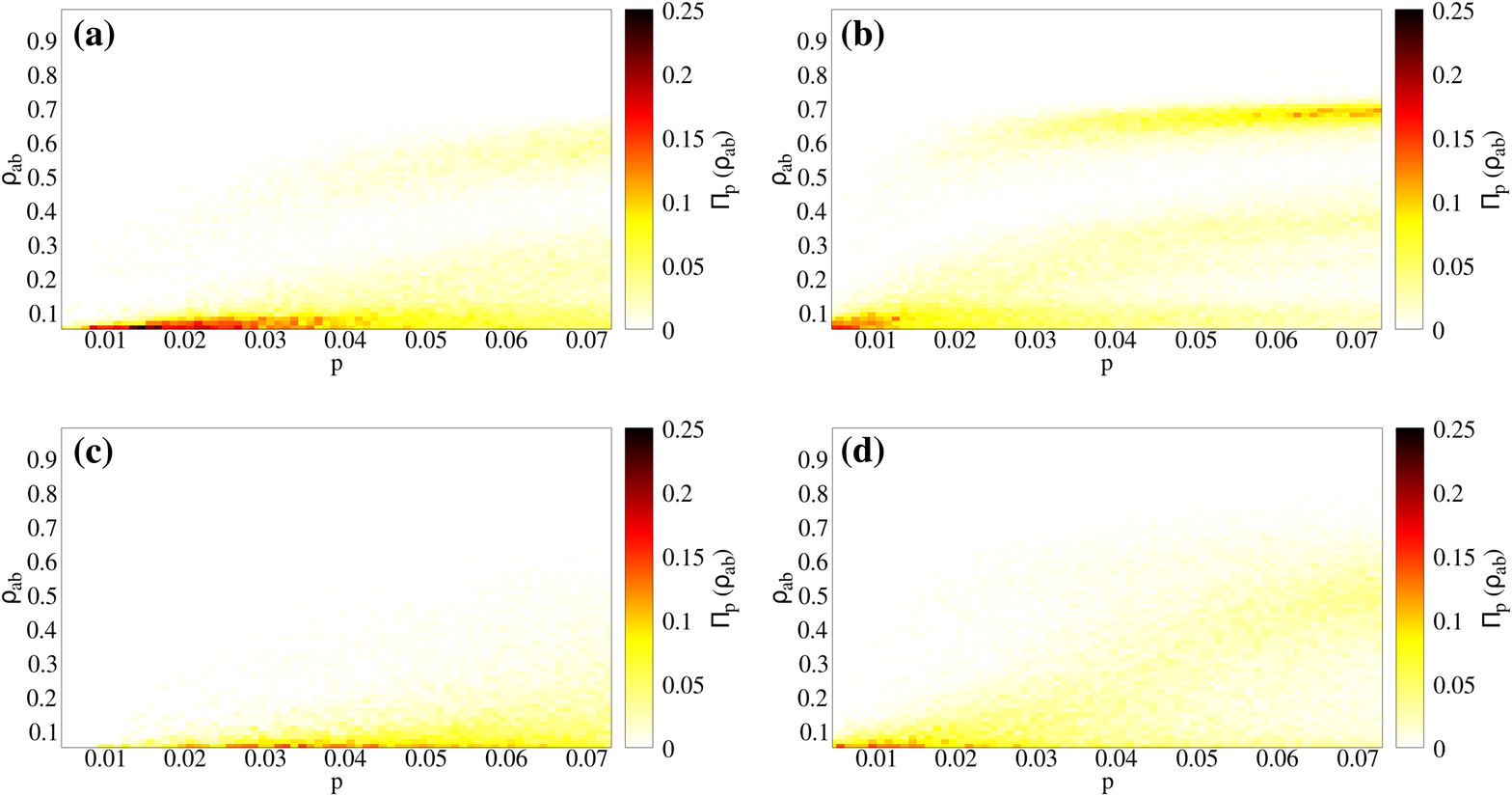}
  \caption{{\bf Spreading process in a temporal contact network.} Prevalence of doubly infected individuals $\rho_{ab}$ as a function of primary infection probability $p$. The colour indicates the fraction of realizations taken which reach the value $\rho_{ab}$ for a given $p$. (a) Independent spreading, with $q=p$, (b) coinfection framework, with $q=1$ , (c) independent spreading ($q=p$) with uncorrelated temporal contact networks with $q$, (d) coinfection framework with an uncorrelated temporal contact networks.}
      \label{fig3}
      \end{figure*}

\begin{figure}[hbt!]
  \includegraphics[width=0.5\textwidth]{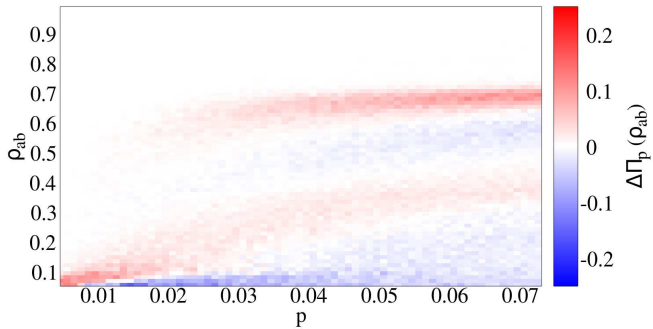}
  \caption{Difference in the prevalence probability $\Delta \Pi_p(\rho_{ab})$ between the cooperative ($q=1$) and independent spreading ($q=p$) cases. In the cooperative case, outbreaks affect a higher fraction of individuals, and happen for a smaller value of the primary infection probability $p$.}
      \label{fig4}
      \end{figure}

We compare two cases: (a) independent spreading ($p=q$), and (b) strong coinfection ($p<q=1$). Both cases lead to similar qualitative results. Interestingly, there are two endemic branches: one growing continuously and another appearing after an abrupt jump (Figs. \ref{fig3}a,b) . However, a quantitative assessment reveals the differences between the independent (Fig. \ref{fig3}a) and cooperative (Fig. \ref{fig3}b) cases. First of all, cooperation makes epidemic threshold smaller, such that outbreaks appear for lower values of $p$ when diseases cooperate, meaning that the risk for the presence of an outbreak is higher, as $p_c$ is smaller. Secondly, outbreaks lead to fractions of doubly infected individuals that are almost a 10\% higher. Finally, for a given value of $p$, the probability of having big outbreaks is higher for the cooperating case, as the highest endemic branch includes a higher density in the cooperative case than in the independent case (Fig. \ref{fig4}).
 
In empirical contact networks, temporal correlations are highly present due to mobility of agents, such that two individuals that are close to each other are more likely to interact soon than when they are in far locations. Moreover, interactions are defined by a time-evolving spatial network, and spatial networks have large values of the clustering coefficient \cite{barthelemy2011spatial}, meaning that if $i$ is connected with $j$ and $j$ is connected with $k$, $i$ is likely to be connected with $k$. In order to study the effect of the temporal structure of our empirical network on coinfection risks, we randomize the sequence of interactions: for every $t$, we randomly choose $C(t)=\sum\limits_{i,j>i}A_{ij}(t)$ contacts from the contact list. In this way, we keep constant both the number of contacts $C(t)$ at every $t$ and the probability for two specific individuals to interact, but we break the correlations mentioned above. Comparing the real with the uncorrelated network for the coinfection and independent cases, the transition from non-endemic to endemic regime has an abrupt jump in the empirical network, while the transition is smooth in the randomized network (Fig. \ref{fig3}c). This highlights the role of temporal correlations as an important factor for the appearance of highly unexpected risks.

\begin{figure}[hbt!]
  \includegraphics[width=0.5\textwidth]{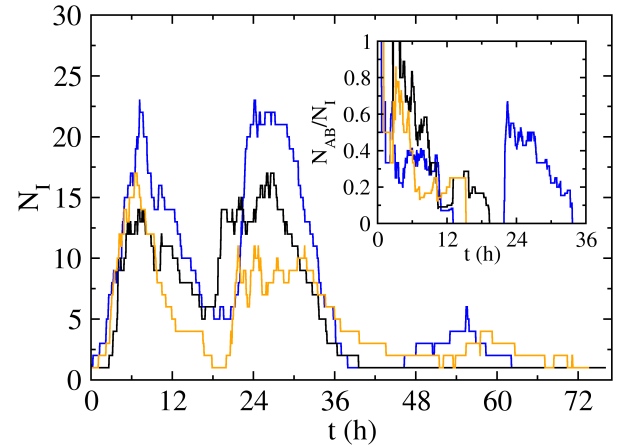}
  \caption{Time evolution of the number of infected individuals $N_I$ for three long-lived realizations with $p=0.06$ and $q=1$, leading to a final number of infected of 56 (blue curve), 54 (black) and 52 (orange) individuals, while the final number of doubly infected is 50 (blue curve), 21 (black) and 14 (orange) individuals. Inset: time evolution of the fraction doubly active individuals $N_{AB}$ amongst the total active individuals $N_I$. }
      \label{fig5}
      \end{figure}

The strong clustering in the high activity periods, which disappears when we break the temporal correlations, makes both infections spread together. In fact, the number of doubly active individuals $N_{AB} (t)$ (\textit{i.e.}, in state $AB$) is a high fraction of the total number of active individuals $N_{I}(t)=N_A(t)+N_B(t)+N_{AB}(t)+N_{Ab}(t)+N_{aB}(t)$ in our system (Fig. \ref{fig5} and inset): even if diseases initially spread following different paths, the strong clustering makes them meet after short paths. However, the low activity periods (\textit{i.e.}, night) are dominated by stochastic recovery processes due to the absence of contacts, leading to a decrease in the number of both active and doubly active individuals. If both infections are able to survive the night ($N_I\neq0$), they may remain active in different individuals ($N_{AB}=0$, blue and orange curves in inset of Fig. \ref{fig5}). In next high activity period, the infections will initially spread independently, but if they meet (blue curve on Fig. \ref{fig5}), they will continue spreading together, reinforced by the cooperative interaction in the coinfection case ($q=1$). As a conclusion, when the spreading process starts from a doubly infected seed on a temporal cluster (high activity period) in our empirical network, the temporal correlations make infections spread together, but in the low activity periods, due to the stochasticity of the recovering process, infections can remain active in the same individuals (state $AB$), recover totally (state $ab$) or partially (states $Ab$ and $aB$). If they continue together (\textit{i.e.}, some $AB$ states remain in the system), or they are able to meet, they can induce similar processes, leading to a high fraction of doubly recovered individuals at the end of the dynamics. In the opposite case, they will spread independently and the fraction of doubly recovered individuals at the end of the dynamics will be lower, just including those that were doubly infected in the first activity period, with a prevalence $\rho_{ab}$ that will be smaller. Those two cases lead to two endemic branches: the upper branch, happening when diseases meet after the low activity period, and the lower branch, growing more slowly, in the opposite case. The process of infections spreading together is reinforced with a higher probability of getting a secondary disease ($q>p$), explaining why the highest endemic branch is more probable and appears for a lower value of the control parameter $p$ in the cooperative case ($q=1$, Fig. \ref{fig3}b), in contrast to the non-interacting case ($q=p$, Fig. \ref{fig3}a).

Previous results of this coinfection model on static networks reported the role of dimensionality, such that low-dimensional lattices, with high local clustering,  lead to continuous transitions, while discontinous transitions appear in lattices with dimension higher than three, which have relatively lower values of local clustering \cite{cai2015avalanche,grassberger2016phase}. The topology explained these results, as diseases spread independently through long paths in the network and, after infecting a macroscopic fraction of individuals, they meet, coinfecting those individuals due to the high value of $q$. This leads to two possible solutions for $\rho_{ab}$ in the region above, but close to the epidemic threshold: if both diseases are not able to meet, $\rho_{ab}\approx 0$, while if they meet, $\rho_{ab}$ has a high value. In fact, finite system size effects may hide the broad jumps, appearing even in networks with broad degree distributions \cite{cui2017mutually}. In our case, similar mechanisms, including the influence of the temporal connectivity pattern, explain how the cooperative interaction reinforces the upper endemic branch: infections spread together in the temporal clusters associated with high activity periods, while the night valley of activity and the stochastic recovery allow diseases to separate and spread independently, leading to macroscopic coinfection effects in $\rho_{ab}$ for the cases in which diseases meet again.
 
After describing the effects of cooperation between infections and temporal correlations on this spreading process, we focus on a specific example. Given that the basic reproductive number of a special influenza strain is $R_0=2$ \cite{mills2004transmissibility}, we estimate the transition probability considering that, for a sufficiently big time window, the network is described by a single giant cluster, and we set a low recovery probability such that an infected individual is able to contact most of the individuals in the system before becoming recovered, leading to $p=\frac{R_0}{N}\approx 0.03$. For this value of $p$, in the independent infections case the system is below the epidemic threshold, while it is above it for the coinfection case, where the outbreaks leads to $\rho_{ab}=0.69$ of the population at most, with the peak for $\Pi_{p=0.03}(\rho_{ab})$ in the highest endemic branch at $\rho_{ab}=0.63$. The coinfection case for the uncorrelated network, for this $p$, is above the epidemic threshold, but leading to lower outbreak prevalences ($\rho_{ab}=0.61$ at most, peak for $\Pi_{p=0.03}(\rho_{ab})$ in $\rho_{ab}=0.13$), as the prevalence grows continuously with $p$ (Fig. \ref{fig3}).

\section{Summary and Conclusion}
Mathematical modelling of disease spreading can help understanding how diseases spread in networks and guide policy makers for better strategies to avoid large outbreaks. To bridge the gap, we applied a coinfection model to an empirical network. Then we estimated the risk of coinfection with three indicators: 1) the epidemic threshold that determines whether there is an outbreak or not, 2) the fraction of the population affected by the outbreak and 3) the probability for having a big outbreak.
 
The combination of cooperation with temporal variables for determining the network of interactions leads to a complex dynamics. In order to determine the role of each of these features, we have splitted our analysis in four cases: (a) cooperative interaction between the infections with temporal correlations, (b) no interaction between the diseases, keeping temporal correlations, (c) cooperative interaction between the infections without temporal correlations, and (d) no interaction between the infections withouth temporal correlations. Comparing cases (a) and (b), we report a higher risk of double infection for the cooperative case, where the epidemic threshold is smaller, outbreaks lead to a higher fraction of infected individuals and they appear more likely (Fig. \ref{fig4}). However, the presence of the abrupt outbreaks both in non-interacting and coinfection cases suggests that several temporal clusters, separated by low activity periods, are formed in the empirical network and the connection of those clusters leads to those abrupt outbreaks. This is confirmed when breaking the correlations in cases (c) and (d), where even with a coinfection framework, for which abrupt outbreaks have been reported under several conditions \cite{cai2015avalanche,grassberger2016phase,chen2013outbreaks}, the randomized temporal network lead to a continuous transition between disease-free and epidemic states. This means that the specific sequence and the temporal correlations that the empirical network contains are responsible for the abrupt jumps (Fig. \ref{fig3}). 

Considering the correlated network, looking at values of $p$ above the epidemic threshold, there is a higher maximum for the prevalence distribution $\Pi_p(\rho_{ab})$, higher bounds for the prevalence if we compare the cooperative case with the independent spreading. We find clear differences between the correlated and uncorrelated cases, as the first experiences a broad transition, while the second is continuous, leading to higher values both for the maximum prevalence and the prevalence at the peak of $\Pi_p(\rho_{ab})$. Surprisingly, the uncorrelated case, that could stand for a static approach, leads to lower values of final fraction of doubly infected individuals, in contrast with other works that study similar models for one disease spreading in empirical temporal networks \cite{masuda2013predicting} and references inside.

Our theoretical study combined with real data can in general help policy makers to make public health more efficient and save more public budget. In summary, in our assessment of the risks of coinfection, we have shown how the connectivity pattern and the temporal correlations in the contact network presented in a hospital contact network can increase the risks of coinfection outbreaks. Hence, a good policy for avoiding this risks would consist on minimizing the effect of these correlations, for example organizing nursing teams in vertical while keeping horizontal organization for physician, such that the temporal clusters are disconnected, in contrast with the model in which both nursing and physician teams are organized horizontally. Future studies can address how vaccination, length of stays in hospitals and different recovery probability may alter the risks.

\section*{Author Contributions}
All authors contributed equally to this work.

\section*{Funding}
  JPR acknowledges support from the FPU program of MECD (Spain). JPR and VME received funding from SPASIMM (FIS2016-80067-P (AEI/FEDER, UE)). FGh acknowledges support from DFG grant.

\bibliographystyle{unsrtnat}
\bibliography{riskofcoinfection}

\end{document}